\documentclass[12pt]{article}
\usepackage{epsfig}
\usepackage{amsfonts}
\newcommand{\De}{\Delta}

\newcommand{\ga}{\gamma} 

\newcommand{\ka}{\kappa}

\newcommand{\La}{\Lambda}

\newcommand{\Om}{\Omega}

\newcommand{\si}{\sigma}

\newcommand{\be}{\begin{equation}}
\newcommand{\ee}{\end{equation}}
\newcommand{\gsim}{\stackrel{>}{\sim}}
\newcommand{\lsim}{\stackrel{<}{\sim}}
\newcommand{\bea}{\begin{eqnarray}}
\newcommand{\eea}{\end{eqnarray}}
\newcommand{\bean}{\begin{eqnarray*}}
\newcommand{\eean}{\end{eqnarray*}}

\def\spose#1{\hbox to 0pt{#1\hss}}
\def\ltapprox{\mathrel{\spose{\lower 3pt\hbox{$\mathchar"218$}}
\raise 2.0pt\hbox{$\mathchar"13C$}}}
\def\gtapprox{\mathrel{\spose{\lower 3pt\hbox{$\mathchar"218$}}
\raise 2.0pt\hbox{$\mathchar"13E$}}}

\begin{document}

{\pagestyle{empty}

\vspace*{2cm}
\begin{center}
{\Huge\bf Frontiers of the Universe: }\vspace{0.2cm}\\
{\LARGE\bf What do we know, what do we understand?}
\vspace{1cm}\\
{\Large\bf Ruth Durrer }\\
Universit\'e de Gen\`eve\\
D\'epartment de Physique Th\'eorique \\
24, Quai Ernest Anserment, 1211 Gen\`eve 4, Suisse
\vspace{2cm}\\
{\bf Abstract}
\end{center}
In this talk I summarize the topics addressed during the conference. We have 
discussed the most relevant cosmological observations of the last say 10 years
and their implications for our understanding of the Universe.  My finding
throughout this meeting has been that it is amazing how our knowledge of 
cosmological parameters has improved during the last years. It comes as 
a surprise to many, that the simplest inflationary model of adiabatic 
perturbations  agrees well with present data. But it is frustrating that we 
have not made any progress 
in identifying the dark matter, in contrary, we have supplemented it by what we
call the 'dark energy', a gravitationally and esthetically repulsive 
component which is even more mysterious than the dark matter.
We have no  understanding about the origin of these two most abundant 
energy components of the present Universe.

\newpage}

\section{Introduction}
Most of us are convinced that below a certain (very high) temperature and 
density and above a certain (very small) length scale, general relativity 
is the relevant theory for the description of the interaction of matter 
and spacetime.
The Friedmann universe is one of the simplest solutions of Einstein's 
equation and this is probably the main reason these solutions have been 
found and discussed originally by Friedmann. In contrast to Newtonian 
gravity, Einstein's equations do allow for homogeneous and isotropic 
solutions, but they are generically non-static.

Nevertheless, I consider it as a big surprise that the Friedmann universe is 
such a good approximation to the observable universe. As I will argue below,
we have several independent pieces of evidence, that the metric deviations 
from the Friedmann solution are of the order of
\be
 \Psi \simeq  10^{-5} \ll 1~,
\ee
on all cosmological scales from about $0.1$  Mpc up to the Hubble scale, 
$H_0^{-1} \cong 3000h^{-1}$Mpc. The fact that this amplitude
of the gravitational potential is scale independent is equivalent to the
fact that the observed fluctuations obey a Harrison Zel'dovich spectrum.

In the following sections I recollect some observational evidences and 
discuss the most important lessons we learn from them. At several occasions
I shall also formulate some doubts, especially concerning the data analysis.

I shall finish with a brief inventory about what we know and then turn to the 
more interesting even if less well defined question of what we understand.
Clearly, 'understanding' has many levels and probably means something 
different for all of us. In this sense I just present a personal point of view.

I shall conclude with a few words about the direction the field might 
develop into.

\section{Big bang nucleosynthesis and Type~Ia Supernovae}
Big bang nucleosynthesis has been discussed 
 by T. X. Thuan, S. Sarkar, and also A. Lubowich during this 
conference. The agreement of the observed light element abundances with
calculations represents one of the fundamental pillars of big bang cosmology.
There are two cosmological parameters which can be determined by comparing 
nucleosynthesis calculations with observations.
The Helium abundance restricts any additional, non-standard, relativistic 
contribution to the energy density of the Universe. This can be cast in
\be {\De N_\nu} \le 0.2 \qquad {\rm or }\qquad\qquad \Om_{X_{\rm rel} }
	\le 1.16\times 10^{-6}~.
\ee
Here, $X_{\rm rel}$ denotes any 'particle' species which is still 
relativistic at the time of nucleosynthesis ($T\sim 0.1$MeV)  in addition to 
the three species of neutrinos and the photons. Examples are a component of
stochastic gravity waves or some other unknown particles, e.g. sterile
neutrinos, with mass $m_X< 0.1$MeV.

The deuterium and Lithium abundances determine the density of 
baryons in the universe~\cite{Burles},
\be
	\Om_bh^2 =0.02\pm 0.002~.
\ee

Even though we do not understand type Ia supernovae in detail (see 
contributions by J. Niemeyer and C.R. Ghezzi), we have realized 
observationally, that they represent excellent 'modified standard candles'.
The Supernovae type Ia measurements of the past years~\cite{Perl,Riess} 
have led to the 
most surprising result that the present universe is accelerating, in other
words the gravitationally active combination $\rho +3p$ is negative in the 
present universe.  Casting this in terms of an equation of state, 
$p=w\rho$, the data yield $w_{\rm eff} \sim -0.6\pm 0.15$. Or, if we assume
the dark energy of the universe to be in the form of a cosmological 
constant with $w_\La=-1$, which is combined with dark matter which has $w_m=0$,
this can be cast as $\Om_m-0.75\Om_\La =-0.25\pm 0.125$. 
These results have been discussed by K. Schahmaneche 
in this meeting.

Clearly, it 
is still somewhat premature to exclude possibilities like 'grey dust' or
'oscillations of the photons' (see e.g.~\cite{Csaki}). But also the large 
scale structure (LSS) data combined with cosmic microwave background 
anisotropies give strong evidence for a cosmological constant (or 
quintessence). 

The only form of energy which leads to a 'cosmological constant' is 
a constant potential energy (in classical physics) or vacuum energy (in 
quantum physics). But exactly this form of energy has no dynamical 
consequences in all known physical interactions except gravity. Only
{\em differences} in the vacuum energy are dynamically meaningful in  
non-gravitational interactions. Once gravity is taken into account, this
otherwise arbitrary integration constant suddenly becomes dynamically 
important. 

Since in supersymmetric theories the vacuum energy from fermions 
and bosons exactly cancels, the typical value of the vacuum energy expected 
from modern particle physics is
\be\label{lambda}
 \rho_\La = {\La\over 8\pi G} \simeq E^4_{SUSY} \gsim 10^{12}{\rm GeV}^4 
	\simeq 10^{59}\rho_c~.
\ee
Here $E_{SUSY}$ denotes the supersymmetry breaking scale which we know to 
be at least 1TeV or larger. The above mentioned observations find a 
non-vanishing value which is at least $59$ orders of magnitude smaller 
than what one naively expects. This finding represents probably the most 
amazing puzzle in physics. Clearly, we are lacking a basic piece in our
understanding and interpretation of the cosmological constant and/or of 
vacuum energy.

Actually, we can also face this result in a positive way: nobody 
(except perhaps Prof. Burbidge) has expected it! This is the clearest 
evidence that cosmology, which for a long time lived from 'theoretical 
speculations' has become data dominated.

I belive that the accelerating universe and the cosmological constant
represents the biggest puzzle in present physics. There are divers attempts
to address it, but none of them so far could convince a majority (see 
contributions from M. Turner and F. Piazza). More observational constraints 
on this mysterious dark energy are desperately needed. Various 
attempts in this direction have been addressed in the 
contributions by J. Newman, R. Bean and J. Weller.

\section{The cosmic microwave background}
The Universe is permeated by an isotropic thermal radiation at 
$T=(2.728\pm 0.01)$K, the cosmic microwave background (CMB). Anisotropies
in this background radiation (a part from a dipole of $C_1 \sim 10^{-3}$ 
which is due to our motion with respect to the CMB rest frame) are of 
the order of 
\be\label{aniso}
{\De T\over T} \simeq  10^{-5} \simeq \Phi ~.
\ee
The last $\simeq$ indicates that CMB anisotropies are of the same order of 
magnitude as the gravitational potential (Bardeen potential) $\Phi$. 
This isotropy, together with the cosmological
principle (we are not sitting in a special place in the Universe), proves 
that the Universe is close to Friedmann on large scales.

The accurate mapping of CMB anisotropies, especially by the three latest 
experiments Boomerang~\cite{Net}, MAXIMA~\cite{Lee} and DASI~\cite{Hal}
teaches us much more that this: the simplest model of a Harrison Zel'dovich 
(scale invariant) spectrum  of purely adiabatic scalar fluctuations with
reasonable cosmological parameters fits the data very well (see Fig.~1).

\begin{figure}[ht]
\centerline{\epsfig{file=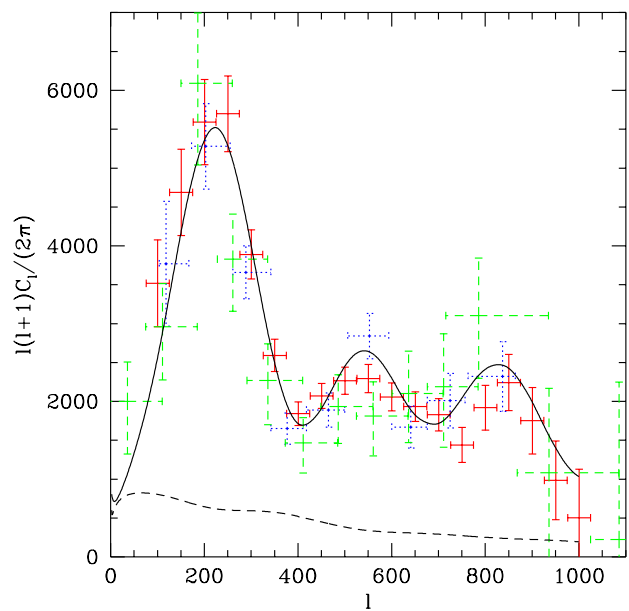, width=3.4in}}
\caption{The CMB anisotropy power spectrum measurements from Boomerang 
(solid,red), MAXIMA (dashed, green) and DASI (dotted,blue) are shown. A typical
inflationary model with scale invariant scalar perturbations is indicated to 
guide the eye (solid line). The CMB anisotropy spectrum for causal global 
defects (texture) is also shown (dashed line).
}
\end{figure}

This result is by no means trivial: the simplest worked out models 
with 'causal' initial perturbations, i.e. initial conditions where all 
correlations on super horizon scales vanish, do not fit the data 
(dashed line in Fig.~1). The reason for this is relatively subtle: these 
models, where fluctuations are induced by topological defects, actually 
also produce a Harrison Zel'dovich 
spectrum of fluctuations. But the fluctuations are not `in phase', 
i.e. at the moment of recombination not all fluctuations of a given 
wavelength are at a maximum/minimum /zero respectively. The peak structure 
which is so characteristic of fluctuations created at a very early initial 
time and which have a fixed amplitude at horizon crossing, is therefore
not expected from topological defects. For topological defects the physical 
mechanism 
which creates the fluctuations shortly after horizon crossing is non-linear. 
The non-linearities imply a coupling of modes which smears out the acoustic 
peaks at least to some extend. Furthermore, tensor and especially vector 
perturbations contribute significantly (more than 50\%) to the CMB 
anisotropies at large scales, leading to a suppression of the purely
scalar acoustic peaks in a COBE normalized spectrum.

This is the main lesson which we have learned so far from these beautiful 
experiments: the cosmological initial fluctuations stem most probably from 
an inflationary phase or some equivalent, which produced fluctuations on 
scales larger than the Hubble horizon. I consider this data as 'evidence for 
inflation' of a completely different quality than the homogeneity and isotropy
of the Universe and the flatness problem. For the latter inflation justifies 
highly improbable initial conditions. In contrary, the acoustic peaks have
been {\em predicted} from inflation. This actually even before the development
of inflation by Doroshkevich, Zeldovich and Sunyaev (1978)~\cite{DZS}. 
These authors have used an initial spectrum of fluctuations with correlations
on super Hubble scales which can only be generated during 
 an inflation-like process. Here we use the term 'inflation'
in its most generic sense as a period where the co-moving size of the Hubble 
scale is increasing.

Since the details of the acoustic oscillations depend sensitively on 
cosmological parameters like 
\be\label{cospar}
\Om_m,~~~\Om_\La,~~~\Om_bh^2,~~~h,~~~ r=T/S,~~~n_s,~~~n_T,~~~\tau_c
	,~~~\cdots, 
\ee
many of us have used them to determine them. Progress in the determination
of cosmological parameters  
has been reported here by S. Sarkar, J. Silk, M. Tegmark and C. Pryke. 
A.~Melchiorri has discussed to which extend present results will improve 
with new data from the MAP and PLANCK satellite experiments.

Even though this parameter estimation is potentially a very powerful method, 
especially when combined with other datasets, which are desperately needed 
since the CMB alone has important degeneracies, I think we must be careful
not to over-interpret the data. Simply consider the number of 
papers written by us theoreticians (I, myself wrote one of them...) about 
the missing second peak after the first Boomerang-98 results~\cite{deB} 
which have now turned out to be due to a systematic error in the data 
analysis (a miss-estimated beam size). The person who warned me most 
not to over-interpret this marginal feature in the Boomerang power spectrum
was Paolo de Bernardis himself.
As Zwicky~\cite{Zwi} put it: ''If only theorists would know what goes into an 
experimental data point and if only experimenters would know what goes into 
a theoretical calculation, they both would take each other much less 
serious''.

Therefore, I guess we may be confident that the cosmological parameters as 
determined so far seem to lie in the right bulk part, but I do by no means 
take seriously the formal errors quoted in the published papers.

An additional point to be aware of is that these parameter estimations 
always {\em assume} that the correct model is in the family of models 
being parameterized. If one allows for more generic models (e.g. relaxing 
the requirement of adiabaticity) one can find very different 
cosmological parameters which also fit the present CMB data (see Fig.~2).  

\begin{figure}[ht]
\centerline{\epsfig{file=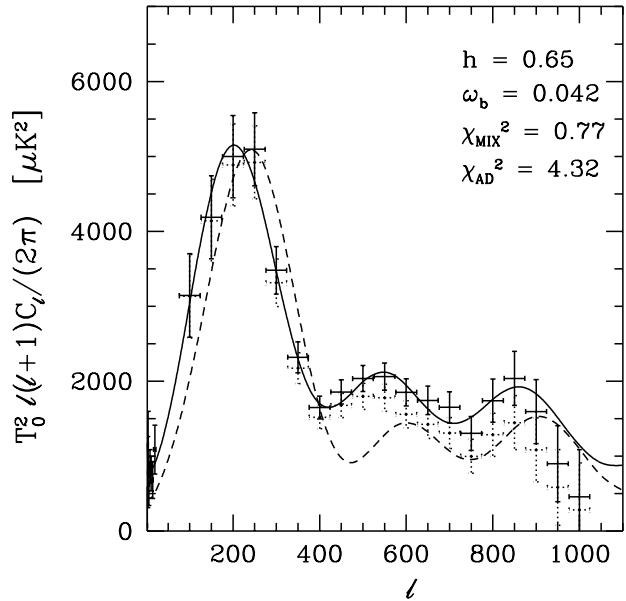, width=3.4in}}
\caption{The CMB anisotropy power spectrum measurements from Boomerang
are compared with flat models with $\Om_bh^2=0.042$ with cosmological 
constant, $\Om_\La=0.7$. The model with mixed isocurvature and adiabatic 
components is a good fit (solid line), while the adiabatic model does not
fit the data with this parameter choice. Figure from Trotta et al.~\cite{iso},
where one can find more details.}
\end{figure}

Much more remarkable than the precision of cosmological parameters obtained 
from CMB measurements so far is their concordance with independent 
determinations from other data like LSS, peculiar velocities, lensing etc.,
under the simplest possible model assumptions. Let us therefore summarize
results from other datasets.

\section{Peculiar velocities}
Apart from the CMB (and lensing), peculiar velocities represent one of the
most straightforward ways to measure the gravitational potential and 
hence the metric perturbations. Unfortunately, they are notoriously difficult 
to measure, since 
\be v_r^{\rm pec} =cz -rH_0  \ee
is the small difference of two large numbers, of which the second one ($r$)
is very difficult to measure. Therefore, one has to consider carefully
how to use this noisy data at its best. A first order of magnitude result is
that peculiar bulk velocities are of the order of 
$v^{\rm pec}(r) \sim 300$km/s on scales of $r\sim 50h^{-1}$Mpc. Using the 
scaling relation from linear perturbation theory,
\be 
v(r) \sim \Phi\times {c^2\over rH_0}, ~\mbox{ this implies }~~  
\Phi \sim 10^{-5}~,
\ee
where $\Phi$ is again the (dimensionless) gravitational potential.

But also on smaller scales, $0.1$Mpc$\lsim r\lsim 1$Mpc, where structures
are virialized and we expect $\Phi \simeq (v/c)^2$, we obtain from the
velocity dispersion in clusters, $v\sim 1000$km/s again $\Phi \sim 10^{-5}$.

Together with the CMB anisotropies on large scales, this  gives 
$\Phi \sim 10^{-5}$ on all cosmologically relevant scales, from the size 
of galaxies up to the Hubble scale, 
$0.1{\rm Mpc}\lsim r\lsim 6000$Mpc.

Clearly, to determine bulk velocity flows, a precise measurement of the
Hubble parameter is desperately needed (see contribution by J. Beckman).
All the results discussed here are not very precise but they are 
nevertheless very interesting and might be telling us something which we 
have overlooked so far...

Finally I want to mention that during the last couple of years, a new 
technique, namely the statistics of pairwise velocities of galaxies has been 
developed and it has been shown that it can lead to a relatively precise
measurement of $\si_8$ and $\Om_m$ as it does not suffer from many of the
problems of bulk velocities~\cite{v12}.

\section{Large scale structure}
New galaxy redshift surveys are being completed or under way. The number of 
published galaxy redshifts is growing very fast. The galaxy redshift catalogs 
are mainly used to determine the galaxy power spectrum which we hope to be 
closely related to the matter power spectrum (see contribution of S. Zaroubi).

Most recently, the 2dF power spectrum has been interpreted to show evidence for
oscillations or at least structure, which indicate a
relatively high ratio $\Om_b/\Om_m\sim 0.15$  (see Ref.~\cite{2dF} and the
contribution by C. Frenk).
It is really amazing to which extent big bang nucleosynthesis, CMB, LSS,
SNIa and also cluster data~\cite{clus} come together and favor 
$h^2\Om_b \sim 0.02$, $\Om_m \sim 0.3$ and $\Om_\La\sim 0.7$.

I admit, however, that I do have reservations to interpret the oscillations
in the 2dF power spectrum as acoustic oscillations of baryons (this concern 
is partially shared with C. Frenk, see his contribution). First of all, 
according to my naive estimate they are not quite at the right position.
Secondly, they could very well mimic some effect from a finite size filter,
the Fourier transform of a top-hat filter simply gives a Bessel function.

I am also a bit reluctant to take seriously the interpretation of the 
'bend in the power spectrum'. Since the mean density of the universe is set
equal to the mean density of  galaxies in the catalog at hand, the power 
spectrum at the scale of the survey, $L$, vanishes by construction 
$P(k=2\pi/L)\equiv 0$. It is therefore not surprising that all observed
galaxy power spectra show a bend in the last few points, {\em i.e.} for 
the smallest values of $k$.

Furthermore, a volume limited sample out to the largest scale
$k\sim 0.02h/$Mpc  can be estimated to contain roughly 1\% of all the 
galaxies in the 2dF catalog leading to $N\sim 1000$. But the Poisson noise 
due to this finite number is just of the order of the square of the 
fluctuation amplitude, $\De^2(k=0.02h/{\rm Mpc})= k^3P(k) \sim 10^{-3}\sim 1/N$.

Therefore, and also due to a 'tradition of sloppiness' in the 
statistical treatment of galaxy redshift data, I am not completely 
convinced that these data is correctly interpreted in the present literature.

\section{Weak lensing}
The deflection of light is fully determined by the gravitational potential, 
{\em i.e.} by the clustering properties of matter. The statistical 
distribution of the shear, the direction of elongation of galaxies, measures 
the gravitational field along the line of sight. We define the
 amplification matrix $A$, 
\be 
\left( \begin{array}{c} x \\ y \end{array}\right)_{\rm image} = A
\left( \begin{array}{c} x \\ y \end{array}\right)_{\rm object}
  \mbox{ with }~~~ A=
\left( \begin{array}{cc} \mu + \ga_1 & \ga_2  \\ \ga_2 & \mu-\ga_1 
\end{array}\right)~.
\ee
The expectation value of $\ga$ is related to the convergence power spectrum
$P_\ka$ via
\be
 \langle\ga^2\rangle ={2\over \pi\theta_c^2} \int_0^\infty{dk\over k}
	P_\ka(k)J_1^2(k\theta_c)~,
\ee
where $J_1$ is the Bessel function of order 1. More details can be found 
in the contribution by F. Bernardeau and in Ref.~\cite{len2}.
 Observations of the shear of galaxies finally lead to constraints in the
$\Om_m$, $\si_8$ plane. For present surveys, the best results are those 
of Ref.~\cite{len2} shown in Fig.~3, but this method is just coming up and
will hopefully lead to improved results, once the errors are better 
under control.
\begin{figure}[ht]
\centerline{\epsfig{file=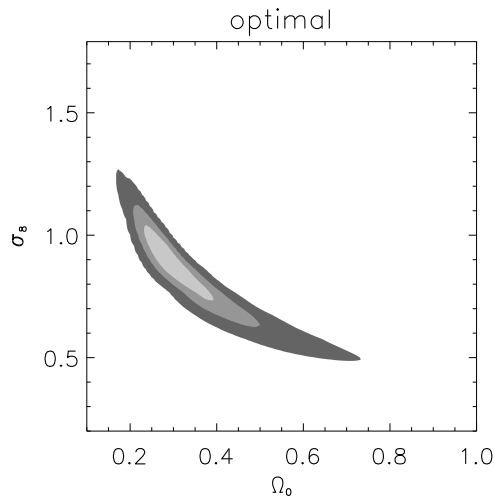, width=3.4in}}
\caption{The 1,2 and 3 $\sigma$ contours for $\Om_m$ and $\si_8$ from the 
weak lensing analysis of the VIRMOS deep imaging survey.
Figure from Werbaeke et al.~\cite{len2},
where one can find more details.}
\end{figure}

\section{Conclusion}
We have seen in this conference that cosmology is harvesting an enormous 
amount of data leading to ever better measurements of the cosmological
parameters. At present it seems that the Universe is flat and 
$\La$-dominated with purely adiabatic scalar perturbations which are 
responsible at the same time for the CMB anisotropies and for large scale 
structure. Astrophysical observations presently undergo amazing breakthroughs
in almost all wavebands of the electromagnetic spectrum. This has 
been impressively demonstrated in the talks by A. Quirrenbach, A. Omont, 
A. Watson, H. Meyer, C. Cesarsky and A. Melchiorri. Also the hope that 
we may finally be able to detect gravitational radiation directly has
become realistic (see contributions by T. Damour and M. Huber).
Observational cosmology is clearly in its most exciting phase, new 
discoveries and challenges for theorists are to be expected every day.

The main parameters of the cosmological model may very well be determined with
sufficient accuracy within a couple of years. Then we will know the parameters
of the Universe we are living in.

But do we understand cosmology? I think there we are still very far from 
our goal: of the two main constituents of the present Universe, we have no
information other than their cosmological activity. Despite intensive
searches (see contributions from T. Sumner, L. Baudis and others), we have 
not observed the particle giving rise to $\Om_m$ yet. There are many
dark matter candidates (see contribution from S. Scopel), but clearly not
all of them can be realized in nature.
Concerning the dominant term in the cosmic density, $\Om_\La$, the situation 
is even worse. There we have no convincing theory giving a result even in 
the right bulk park. Interesting ideas are being developed
 (see {\em e.g.} the contribution by T. Padmanabhan), but so far the problem 
remains unsolved.

Also inflation, the 'explanation' of the flatness and homogeneity of the 
universe, and of the adiabatic initial fluctuations, is more on the level
of a paradigm than of a theory of the early universe, motivated by 
high energy physics (see contribution by S. Dodelson). 

These are, as I see it, the main problems which concern the early universe.
Most probably, they cannot be solved without simultaneous progress in the
theory of high energy physics. 
They may actually provide our only clues to the physics at very high energies,
like string theory, which are probably not accessible also in future
accelerators (see the contributions from G. Veneziano and J. Magueijo).
Cosmology may turn out as the ultimate tool to study the fundamental laws 
of physics.

At the
redshifts $10^{10} \gsim z\gsim 10$, {\em i.e.} from about $T\sim 1$MeV until 
cosmic structures become non-linear and the first objects form, we 
understand the cosmic evolution rather well. We can provide good,
successful calculations of primordial nucleosynthesis and recombination, as
well as integrate the linear perturbation equations to determine CMB 
anisotropies. These calculations are in good agreement with observations
and, together with other measurements, they allow us to determine the 
cosmological parameters. 

At low redshift, $z<10$, when structure develops, radiation processes and 
chemical evolution take place together with gravitational clustering. The 
physics becomes complex. Many different effects have to be taken into 
account simultaneously, if we want 
to obtain a consistent picture. Complex numerical simulations combining 
gravity with hydrodynamics are
presently being developed (see contribution from T. Abel). But also 
new observations focusing on small scales in different bands of the 
electromagnetic spectrum are underway (see
contributions from S. Bridle, I. Lehmann, A. Bunker, E. de Gouveia Dal Pino 
N. Gruel and others).

I believe that, within a couple years when the important work of determining 
cosmological parameters to good accuracy will have been solved, there will 
remain two major directions for cosmological research:
\begin{itemize}
\item $z\gg 10^{10}$ Early universe, towards the fundamental laws of physics.
\item $z\lsim 10$~~~ The formation of galaxies and stars, complexity.
\vspace*{1.5cm}
\end{itemize}
\vspace{0.5cm}

\noindent {\bf Acknowledgment:} I thank the participants of this 
'Rencontre de Blois' and Filippo Vernizzi for interesting and stimulating 
discussions. My participation at the conference was supported by the Swiss 
National Science Foundation.

\end{document}